\documentclass[a4paper, amsfonts, amssymb, amsmath, reprint, showkeys, nofootinbib, twoside]{revtex4-1}
\usepackage{graphicx} 
\usepackage{dcolumn} 
\usepackage{bm}
\usepackage{hyperref} 
\usepackage[mathlines]{lineno}
\begin{document}
\preprint{APS/123-QED}
\title{Coulomb Form Factors of $^{27}Al$ and $^{31}P$ Nuclei Using Coulomb valance Tassie model and Bohr-Mottelson Collective Models with Different Potentials}
\author{Khalid S. Jassim}
\affiliation{Department of Physics, College of Education for pure Science, University of Babylon, PO Box 4, Hilla-Babylon, IRAQ} 
\affiliation{khalidsj@uobabylon.edu.iq}
\author{Ranya Mahmoud Mohammed}
\affiliation{Al-Zahraa University for Women, College of Health and Medical Technology, Department of Radiological Technology, Kerbala, Iraq.} 
\date{\today}
\begin{abstract}
The longitudinal Coulomb C0, C2 and C4 form factors with core-polarization effects have been studied using shell model calculations for ${3/2}^+_1$ state with excitation energy of 1.069 MeV, ${5/2}^+_2$ state with excitation energy of 2.706 MeV and ${7/2}^+_1$ state with excitation energy of 2.304 MeV state in $^{27}Al$ nucleus and ${1/2}$ state in $^{31}P$ nucleus. The two-body effective interaction Wildenthal and universal sd-shell interaction A (USDA) are used for sd-shell orbits. The Coulomb valance Tassie model (CVTM) and Bohr-Mottelson (BM) collective model have been used to calculted the Core-polarization effects. Three potentials are Harmonic Oscillator (HO) potential, Wood-Saxon (WS) and SKX potential,which have been used to calculated the wave functions of radial single particle matrix elements. The results for these potentials are compared with final update experimental data. The calculations with inclusions of core-polarization effects give a good agreements of experimental data.  
\end{abstract}
\maketitle
\section{Introduction:}
In nuclear physics calculations, shell model considered as the suitable theoretical model for finding and understanding the properties of different nuclei. It can be also considered as a basis for calculations that are much more complex and complete such as higher shell state nuclei. One can say that the shell model is widely used as a suitable model for low-lying excitation in a variety of nuclei. The shell-model is assumed that all the nuclei can separate into a inert core part and a valence particles part, and describe the interaction between them. 
The high-energy electron scattering from the nucleus is very useful as a means of showing the nuclear structure \cite{1}.The electron scattering is the best method to explain the distribution of charge in nucleus, in addition to the fact that the electron particle points consequently the electron that can readily accelerated \cite{2}.
The shape of the nucleus is determine by Electron scattering form factor. It depends on current, magnetization distribution and the charge in the target nucleus. Experimentally, form factors can be found as a function of the momentum transfer (q) by knowing the energies of the incident and scattered electrons and the scattering angle \cite{3}. There are two types of form factor are: the first is longitudinal (coulomb) form factor, which represents the Fourier transform of the charge density and provide all information about the nuclear charge distribution. The second kind is transverse form factor, which is represented the Fourier transform of the current density. The second type contain all the information about the magnetization distributions and nuclear current\cite{4}. According to the parity selection rules, the transverse form factor can be sorted into transverse electric and magnetic form factor.
All the types the interaction between inert core and valance particles causes excite the core. This process, known as core excitation, this process can be described as a polarization of the core by one of the valence particles and gives hole in the inert core, which is called core-polarization (CP) effects. The universal sd-shell interaction (USD) Hamiltonian \cite{3} provided realistic sd-shell ($1d_{5/2}$, $1d_{3/2}$) wave functions for use in nuclear structure models, nuclear spectroscopy, and nuclear astrophysics for over two decades. It is also an important part of the Hamiltonian used for the p-sd \cite{5} and sd-pf \cite{6,7,8} model spaces\cite{9}. Coulomb C2 and C4 form factors of $^{18}O$, $^{20,22}Ne$ nuclei using Bohr- Mottelson collective (BM) model have been studied by Ajeel and et. el.\cite{10}. Elastic and inelastic C2 form factors of nickel isotopes and two-body correlations have been studied by Abbas and et. al. \cite{11}. Li, Xin and et. al. have been made Comparative studies on nuclear elastic magnetic form factors between the relativistic and non-relativistic mean-field approaches\cite{12}. Khalid S. Jassim and et. .al. have been worked some theoretical paper of electron scattering form factors using shell model calculations of many light and medium nuclei.\cite{13,14,15,16}.

\section{Theory:}
CVTM one of the important model, which is used to calculate the core-polarization effect. Nushellx@MUS is a modeling of more elasticity and modification that allows a non-uniform mass and charge density distribution\cite{15}. The polarization of core (CP) charge density in CVTM model depends on the ground state charge density of the nucleus. The ground state charge density is expressed in terms of the two-body charge density for all occupied shells including the core. Based on the collective modes of the nuclei, the Tassie shape CP transition density is given by\cite{17},
\begin{eqnarray}
\rho_{Jt_z}^{core}\ (i,f,r) =1/_2C (1+\tau_z) r^{J-1}\frac{d\rho_0\ (i,f,r)}{dr}
\end{eqnarray}
where C is a constant of the proportionality and $ \rho_0 $ is the charge density distribution for the ground state two-body, which is given \cite{18}

\begin{equation}
\rho_0 =\langle \psi \|\hat{\rho}_{eff}^{(2)}\ (\vec{r})  \|  \psi\rangle\ =\displaystyle\sum_{i<j} \langle ij \|\hat{\rho}_{eff}^{(2)}\ (\vec{r})  \|  ij\rangle\ - \langle ij \|\hat{\rho}_{eff}^{(2)}\ (\vec{r})  \|  ji\rangle\ ,
\end{equation}
where
\begin{eqnarray}
 \hat{\rho}_{eff}^{(2)} (\vec{r}) = \frac{1}{2(A-1)}  \displaystyle\sum_{i<j} \{ \delta (\vec{r}-\vec{r_i}) \\ \nonumber
 + \delta (\vec{r}-\vec{r_j}) \} f(r_{ij}) 
 \end{eqnarray}
where i and j are all the required quantum numbers, i.e., the functions $f(r_{ij})$ are the two body short range correlation (SRC). In this work, a simple model form of short range correlation has been adopted, i.e.\cite{19},
\begin{equation}
\ f(r_{ij})=1- exp[-\beta (r_{ij}-r_c)^2]
\end{equation}

where $r_c$ is the suitable hard core radius and $ \beta $ is a parameter of the correlation. 
\begin{widetext}
The Coulomb form factor for this model becomes\cite{18}:
\begin{equation}
F^L_J(q)= \sqrt\frac{4\pi}{2J_i+1} \frac{1}{z}\left[\int_0^\infty \mathrm r^2 J_j(qr) \rho_{jt_z}^{m_s} (i,f,r) dr + C\int_0^ \infty \mathrm  j_J(qr) r^{J+1} \frac{d\rho_o (i,f,r)}{dr}\right] F_{cm}(q) F_{fc}(q)
\label{eq:wideeq}
\end{equation}
\end{widetext}
but
\begin{eqnarray}
\int_0^ \infty \mathrm j_J(qr) r^{J+1}\frac{d\rho_o (i,f,r)}{dr}\\ \nonumber
= \int_0^ \infty \mathrm \ \frac{d}{dr} [j_J(qr) r^{J+1}\rho_o(i,f,r)]{dr}\\ \nonumber
-\int_0^ \infty \mathrm \ (J+1) j_J(qr) r^J \rho_o (i,f,r) dr\\ \nonumber
-\int_0^ \infty \mathrm \ \frac{d{j_J}(qr)}{dr} r^{J+1} \rho_o (i,f,r) dr,
\end{eqnarray}
where the first term gives zero contribution, the second and the third term can be combined together as\cite{17},
\begin{equation}
    -q\int_0^ \infty \mathrm \\r^{J+1} \rho_o (i,f,r) [\frac{d}{d(qr)} + \frac{J+1}{qr}] j_J(qr)dr
\end{equation}
from the recursion of the spherical Bessel function\cite{18},
\begin{equation}
      \left[\frac {d}{d(qr)} + \frac{J+1}{qr}\right] j_J(qr)dr=j_{J-1}(qr)
\end{equation}
\begin{eqnarray}
\int_0^ \infty \mathrm  j_J(qr) r^{J+1}\frac{d\rho_o (i,f,r)}{dr}dr\\ \nonumber
-q\int_0^ \infty \mathrm r^{J+1} \rho_o (i,f,r) j_{J-1}(qr)dr
\end{eqnarray}
\begin{widetext}
Therefore, the form factor of eq.(5) takes the form\cite{19}:
\begin{equation}
    F^L_J(q)= \sqrt{\frac{4\pi}{2J_i+1}}   \frac{1}{z}\left[\int_0^\infty \mathrm r^2 J_j(qr) \rho_{jt_z}^{m_s} (i,f,r) dr
    -qC\int_0^ \infty \mathrm r^{J+1} \rho_o (i,f,r) j_{J-1}(qr)\right] F(q)_{cm}F(q)_{fc}
 \label{eq:wideeq}
\end{equation}
\end{widetext}
\
The proportionality constant C can be determined from the form factor evaluated at q=k, i.e. substituting q=k in above equation we obtained\cite{18},
\begin{equation}
    \frac{\int_0^\infty \mathrm r^2 j_j(kr) \rho_{jt_z}^{m_s} (i,f,r) dr-Z F_J^L(k) \sqrt{\frac{2J_i+1}{4\pi}}}
{\int_0^ \infty \mathrm \\r^{J+1} \rho_o (i,f,r)j_{J-1}(qr)dr}
\end{equation}
\section{Results and discussion}
The motivation of the present work is to calculate the effect of core-polarization on electron scattering form factors with different potentials (HO, SKX and WS) and different models (CVTM and BM). Then makes comparisons between these calculations to find the best method and best potentials.

\begin{figure}
\includegraphics[scale=0.4]{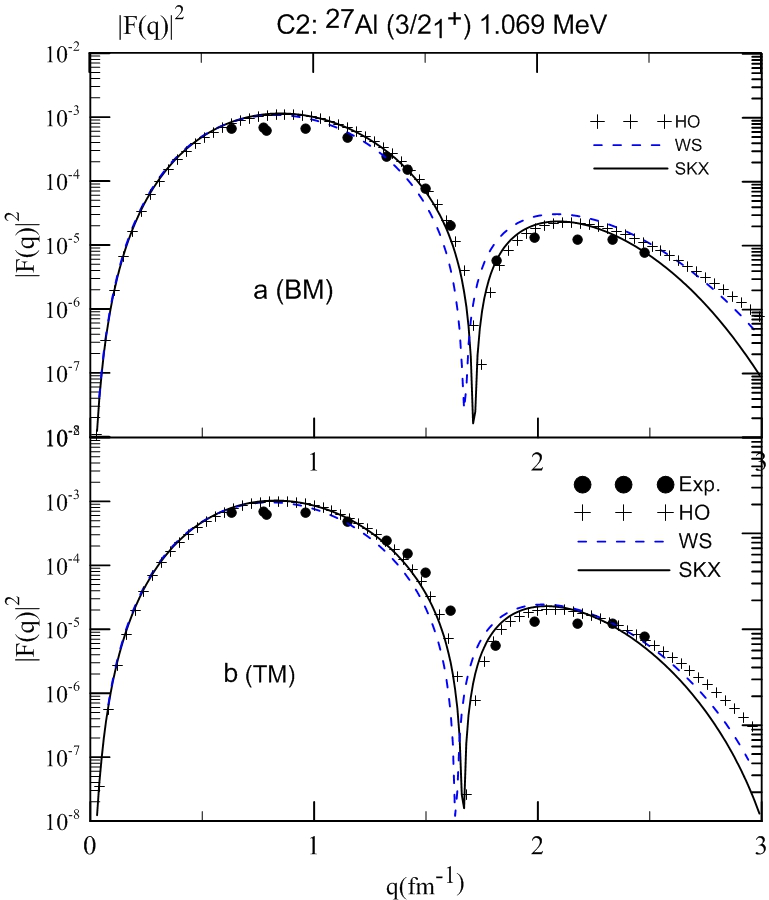}
\caption{\label{fig1:jpg} The longitudinal C2 form factors for the transition of the $3/2_1^+$ (1.069 MeV) state in $^{27}Al$, using BM collective model and Tassie model with three different potential. the experimental data are taken from ref.\cite{21}}
\label{fig1.jpg}
\end{figure}

\begin{figure}
\includegraphics[scale=0.4]{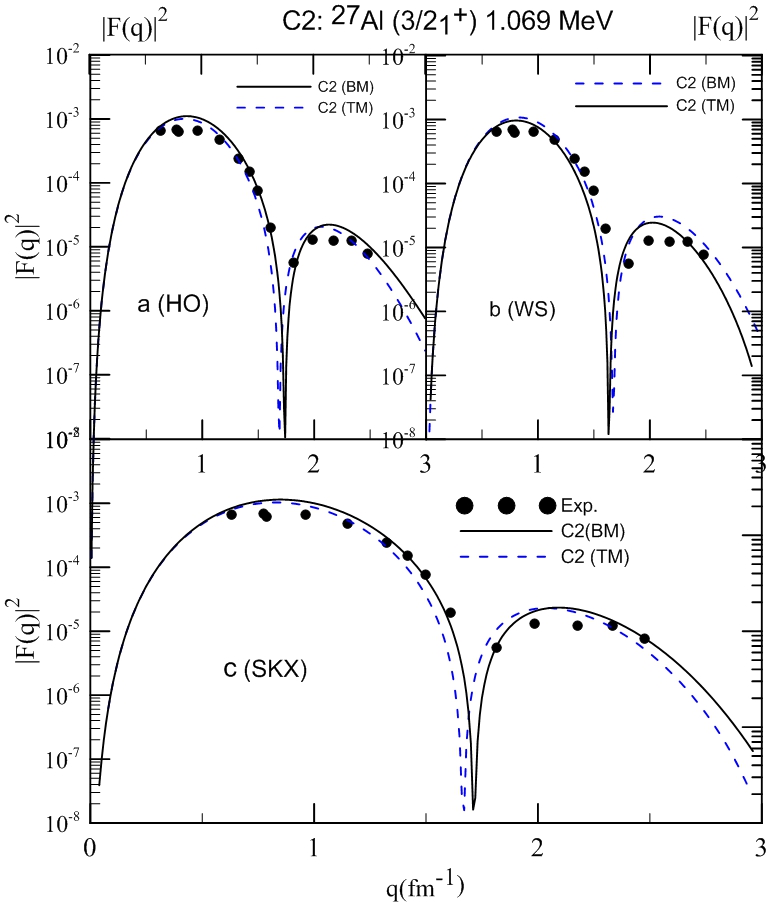}
\caption{\label{fig2:eps}The longitudinal C2 form factors for the transition of the $3/2_1^+$ (1.069 MeV) state in $^{27}Al$.the experimental data are taken from ref.\cite{21}}
\label{fig2.jpg}
\end{figure}

\begin{figure}
\includegraphics[scale=0.4]{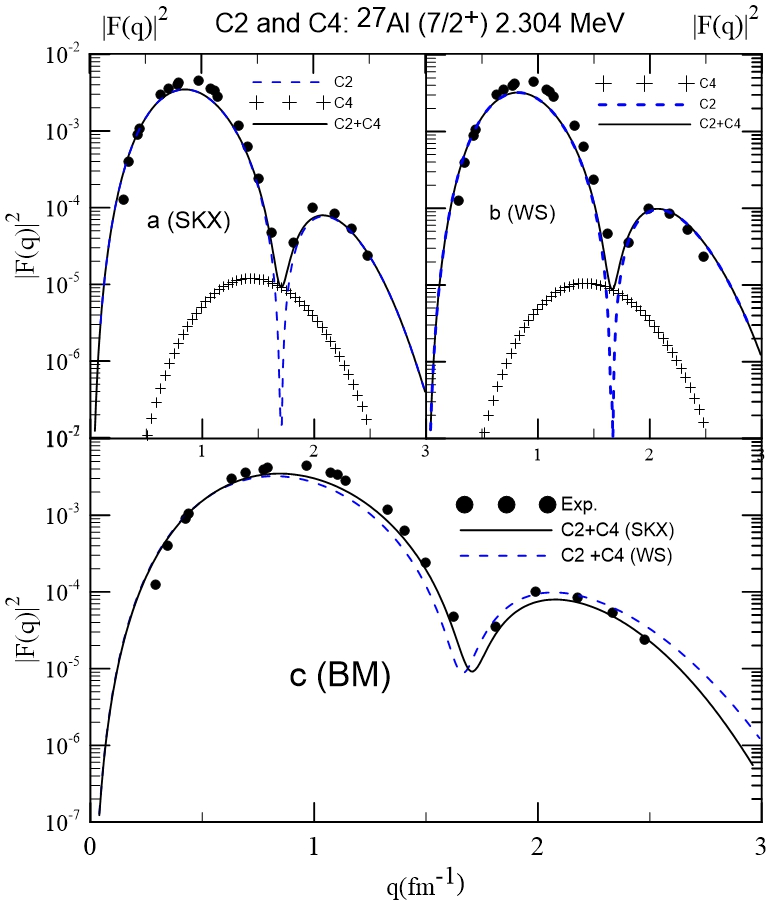}
\caption{\label{fig3:gpj} The longitudinal C2+C4 form factors for the transition of the $7/2^+$ (2.304 MeV) state in $^{27}Al$, using BM collective model with three different potential (HO, WS and Skx). the experimental data are taken from ref.\cite{21}}
\label{fig3.jpg}
\end{figure}

\begin{figure}
\includegraphics[scale=0.4]{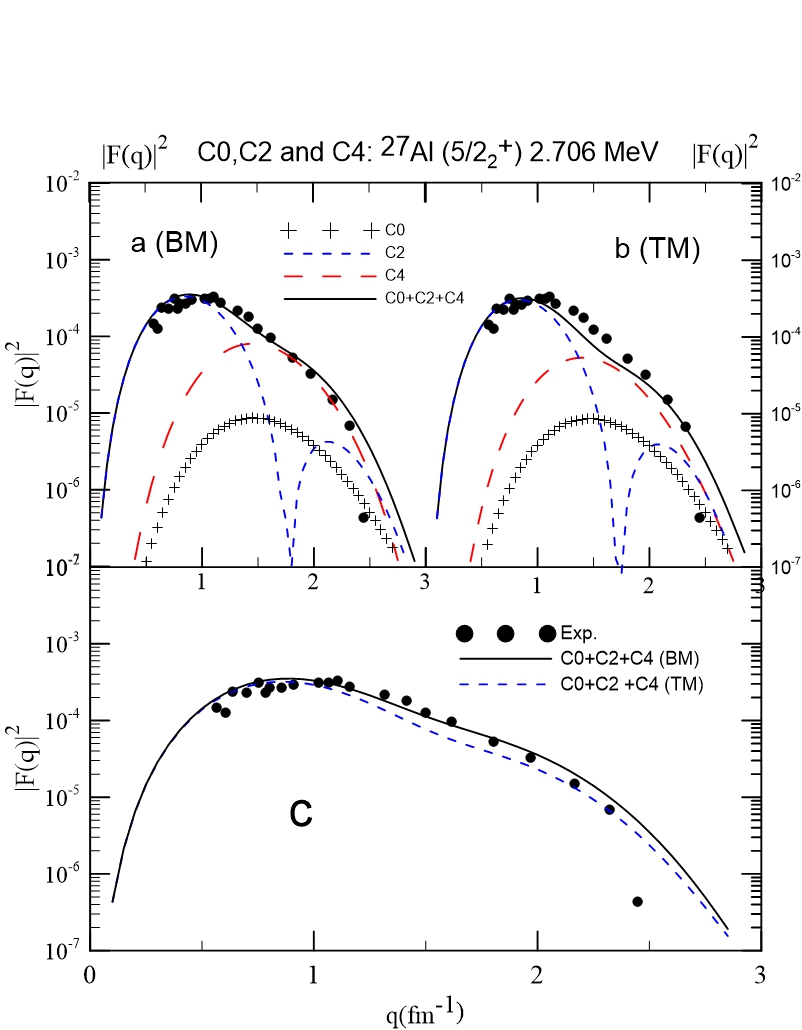}
\caption{\label{fig4:eps}The longitudinal C2 form factors for the transition of the $5/2_2^+$ (2.706 MeV) state in $^{27}Al$, using BM collective model and Tassie model with Skx potential. The experimental data are taken from ref.\cite{21}}
\label{fig4.jpg}
\end{figure}

\begin{figure}
\includegraphics[scale=0.4]{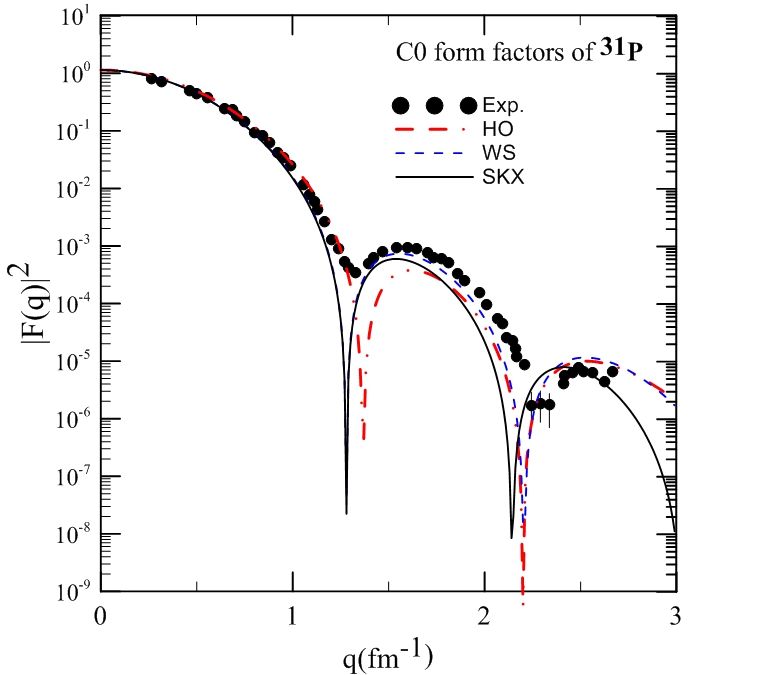}
\caption{\label{fig5:jpg}The longitudinal C0 form factors for the $^{31}P$, using BM collective model, with HO, WS and SKX potential.The experimental data are taken from ref.\cite{22}}
\label{fig5.jpg}
\end{figure}

Calculations are showed the Coulomb C2 in $^{27}Al$ with excitation energies 1.069 MeV, 2.304 MeV and 2.706 MeV for ${3/2}_1^+$, ${7/2}^+$  and ${5/2}_2^+$, respectively, and C0 for $^{31}P$ with excitation energies 0.0 MeV.These calculations are performed using USDA effective interaction for sd-shell model to generate the OBDM elements. These calculations are performed using the shell model Nushellx@MUS code\cite{20}. The wave functions for single-particle are these of the Harmonic Oscillator (HO), Wood Saxons (WS) and Skyrman (SKX) potentials. According to the sd-shell model concepts, it is described as an inert core of $^{16}O$ plus 11, 15 nucleons for $^{27}Al$ and $^{31}P$, respectively, which distributed over sd-shell. The comparison between the theoretical and experimental C0 and C2 form factors.
The longitudinal C2 form factors for ${3/2}_1^+$ (1.069 MeV) state in $^{27}Al$ nucleus calculated three potentials: HO, WS and SKX on sd-shell model wave function, which are shown in figs.\ref{fig1.jpg}.The CP effects with Tassie model and BM collective model are shown in \ref{fig1.jpg}, in this figure which represent the relation between the C2 form factors as a function of momentum transfer q. The panel (a) and (b) notice the electron scattering with BM collective model and CVTM model, respectively, while in fig. \ref{fig2.jpg} with panel a, b and c are show the comparison between BM collective model (solid curves) and CVTM model (dashed curves) with three different potentials HO, WS and SKX, respectively.
the best fit of form factors comparing with experimental data at the skx potential, this potential gives agreement in all momentum transfer region, comparing with other  (HO and WS) potentials.The others potentials, so give agreement in the first and second maximum regions.  
Figs.\ref{fig3.jpg} are show the calculated total C2+C4 form factors with the inclusion of CP effects by BM collective model for three potentials (Skx, WS) as shown in banal (a) and (b), respectively. The compassion between C2+C4 form factor with BM model for SKx and WS as shown in banal C by the solid curve for SKX potential and the Dashed curves for WS potential. The theoretical data are in a good agreement for in the first minimum region and second maximum region momentum transfer. 
In fig.\ref{fig4.jpg}, the calculations of longitudinal C0+C2+C4 using CVTM and BM collective model with Skx potential. all the theoretical results are give agreement in all momentum transfer region, but the results with BM collective model are more closer to the experimental data in the momentum transfer region between (1.2-2.0) $fm^{-1}$ comparing with calculations with CVTM. All calculations were performed using Skx wave function.
The C0 form factors for ${1/2}$ with 0.0 MeV state in $^{31}P$ nucleus calculated with three potentials: HO, WS and SKX on sd-shell model wave function, which are shown in Figs.\ref{fig5.jpg}. the results in the first maximum momentum transfer are give a good agreement. in the second region between (1.25-2.1) $fm^{-1}$, the theoretical results with WS potential give a good agreement comparing with HO and Skx, but all potential are give the same behavior comparing with the experimental data.In the third region between (2.1-3.0) $fm^{-1}$, the theoretical results with Skx potential give agreement comparing with HO and Skx results, but all potential are give the same behavior comparing with the experimental data.
In general, the importance of the core polarization effects in some longitudinal components of form factors remains as an open question. The calculations with core polarization effects may be affected by some important parameters such as effective charges (protons and neutrons), the size parameters for HO potential, the type of potentials and the type of effective charge which is used in calculations. However, All models, which is used in the present work give agreements in all momentum transfer region comparing with experimental data.In this work, the calculations are performed without adjusting any parameter.
\section{conclusions}
The coulomb C0, C2 and C4 form factors are calculate with core-polarization effects using CVTM model and BM collective model for ${3/2}^+_1$ with 1.069 MeV ,${5/2}^+_2$ with 2.706 MeV and ${7/2}^+_1$ with 2.304 MeV states in $^{27}Al$ nuclei and ${1/2}$ for $^{31}P$ nuclei. The USDB interactions for sd-shell are used with HO, WS and Skx potentials. The core polarization effects on form factors with using BM collective model are found to be very important in the calculations of the C0, C2 and C4 form factors and gives good agreement over the sd-shell model calculations for the form factors. All results give agreement comparing with experimental data. In this work, the calculations are performed without adjusting any parameter.

\end{document}